# Inflight Radiometric Calibration of New Horizons' Multispectral Visible Imaging Camera (MVIC)

C.J.A. Howett[1], A.H. Parker[1], C.B. Olkin[1], D.C. Reuter[2], K. Ennico[3], W.M Grundy[4], A.L. Graps[5,6], K.P. Harrison[7], H.B. Throop[8], M.W. Buie[1], J.R. Lovering[9], S.B. Porter[1], H.A. Weaver[10], L.A. Young[1], S.A. Stern[1], R.A. Beyer[3], R.P. Binzel[11], B.J. Buratti[12], A.F. Cheng[10], J.C. Cook[1], D.P. Cruikshank[3], C.M. Dalle Ore[3], A.M. Earle[11], D.E. Jennings[2], I.R. Linscott[13], A.W. Lunsford[2], J.Wm. Parker[1], S. Phillippe[14], S. Protopapa[15], E. Quirico[14], P.M. Schenk[16], B. Schmitt[14], K.N. Singer[1], J.R. Spencer[1], J.A. Stansberry[17], C.C.C. Tsang[1], G.E. Weigle II[18], A.J. Verbiscer[19].

1 - Southwest Research Institute, Boulder, CO 80302, USA.

2 - NASA Goddard Space Flight Center, Greenbelt, MD 20771, USA.

3 - NASA Ames Research Center, Space Science Division, Moffett Field, CA 94035, USA

4 - Lowell Observatory, Flagstaff, AZ 86001, USA

5 - Planetary Science Institute, Riga, Latvia

6 - University of Latvia, Riga, Latvia

7 - Consultant, Denver, CO, USA

8- Planetary Science Institute, Mumbai, India

9 - The Breakthrough Institute, Oakland, CA 94612, USA

10 - Johns Hopkins University Applied Physics Laboratory, Laurel, MD 20723, USA

11 - Massachusetts Institute of Technology, Cambridge, MA 02139, USA

12 - NASA Jet Propulsion Laboratory, La Cañada Flintridge, CA 91011, USA

13-Stanford University, Stanford CA 94305, USA

14-Université Grenoble Alpes, CNRS, IPAG, F-38000 Grenoble, France

15-Department of Astronomy, University of Maryland, College Park, MD 20742, USA

16-Lunar and Planetary Institute, Houston, TX 77058, USA

17 - Space Telescope Science Institute, Baltimore, MD 21218, USA

18- Southwest Research Institute, San Antonio, TX 28510, USA

19 - Department of Astronomy, University of Virginia, Charlottesville, VA 22904, USA

**Corresponding Author and their Contact Details:**

C.J.A. Howett

Email: howett@boulder.swri.edu

Telephone Number: +1 720 240 0120

Fax Number: +1 303-546-9687

Address:

1050 Walnut Street, Suite 300

Boulder, Colorado

80302

USA

**Abstract**

We discuss two semi-independent calibration techniques used to determine the in-flight radiometric calibration for the New Horizons' Multi-spectral Visible Imaging Camera (MVIC). The first calibration technique compares the observed stellar flux to modeled values. The difference between the two provides a calibration factor that allows the observed flux to be adjusted to the expected levels for all observations, for each detector. The second calibration technique is a channel-wise relative radiometric calibration for MVIC's blue, near-infrared and methane color channels using observations of Charon and scaling from the red channel stellar calibration. Both calibration techniques produce very similar results (better than 7% agreement), providing strong validation for the techniques used. Since the stellar calibration can be performed without a color target in the field of view and covers all of MVIC's detectors, this calibration was used to provide the radiometric keywords delivered by the New Horizons project to the Planetary Data System (PDS). These keywords allow each observation to be converted from counts to physical units; a description of how these keywords were generated is included. Finally, mitigation techniques adopted for the gain drift observed in the near-infrared detector and one of the panchromatic framing cameras is also discussed.

# 1 Introduction

*1.1 MVIC*

The Multi-spectral Visible Imaging Camera (MVIC) is part of the Ralph instrument on the New Horizons spacecraft. Full details of instrument can be found in Reuter *et al.* (2008), but an overview is provided here for reference. A single substrate holds MVIC's seven independent CCD arrays. Six of these CCDs operate in Time Delay and Integration mode (TDI), and each has 5024x32 pixels. There are two panchromatic TDI arrays with the same wavelength range for redundancy, and four other TDI arrays each with a color filter. The final array is a frame transfer 5024x128 pixel array, primarily designed for optical navigation. The Pan TDI rates can run between 4 and 84 Hz, while the color TDI rates are between 4 and 54 Hz. The frame transfer integration time is 0.25 to 10 seconds. Further details of all these arrays (including typical exposure times) can be found in Tables 1-4.

TDI is a way of building up large format image as the field of view (FOV) is quickly scanned across a scene. It works by syncing the transfer rate between rows to the spacecraft's scan rate, thus the same scene passes through each of the rows before it is read out, effectively increasing the integration time. The frame transfer camera is operated in the more traditional stare mode.

Each of MVIC's detectors has its own response ($\chi$) as a function of wavelength ($\lambda$), which is calculated according to Equation 1,

$$\chi(\lambda) = Q_e(\lambda)F_t(\lambda)B_S(\lambda)A_l^3(\lambda), \qquad \textit{[Equation 1]}$$

where $Q_e$ is the quantum efficiency of the detector, $F_t$ is the filter transmission, $B_S$ is the beam splitter reflectance and $A_l$ is the mirror reflectance (it's cubed because there are three mirrors in the system). Reuter *et al.* (2008) provided MVIC's quantum efficiencies and filter transmissions measured before launch. Figure 1 shows these values, along with the measured beam splitter reflectance and mirror reflectance. The figure shows that when these values are combined to give a responsivity many of the color detectors have a larger response than either of the panchromatic detectors at some wavelengths. The ground-based spectral sampling of the detector's $Q_e$ was very coarse (~50 nm), so it is possible these curves are not accurately characterized. A minor correction was made to restrict the peak of all the color detector's $Q_e$ to that of the panchromatic detector, as it was deemed unlikely that their response would be higher than that of the clear filter given the identical nature of all the CCDs (Reuter *et al.*, 2008). This correction has an almost negligible effect on the final result, as their absolute values will change as a result of this calibration. The new responsivities used in this work are also shown in Figure 1. Using this responsivity the effective (or pivot) wavelength $\lambda_p$ for a given filter is calculated according to Equation 2, where $\chi$ and $\lambda$ are defined as above (Laidler *et al.*, 2005).

$$\lambda_p = \frac{\int \lambda \chi(\lambda) d\lambda}{\int (\chi(\lambda)/\lambda) d\lambda} \qquad \textit{[Equation 2]}$$

*1.2 Calibration Outline*

This paper presents two parallel and semi-independent MVIC inflight radiometric calibration processes and one process for "bootstrapping" a correction for a gain drift identified in one of MVIC's two redundant sets of readout electronics. One radiometric calibration process, based on deriving system throughput corrections using photometry of calibration stars, is planned to become the long-term standard for MVIC calibration. The second radiometric calibration process is based on deriving corrections to the mean observed color ratios of Charon in order to match color ratios measured by the Hubble Space Telescope. It was developed for the Blue, Red, NIR, and CH4 MVIC channels on approach to the Pluto system to expedite a well-understood interim radiometric calibration solution in order to enable certain science observations (e.g., mapping the CH4 ice distribution across Pluto using Red, NIR, and CH4 MVIC imagery; see Grundy et al. 2016). This Charon-based calibration was utilized by a number of the early New Horizons-based science papers (e.g., Grundy et al. 2016, Weaver et al. 2016). The following sections outline these calibration procedures, their results, and how the results of each procedure compare.

**2 Stellar Calibration**

*2.1 Calibration star observations*

Each year during the 9.5-year cruise to the Pluto-Charon system the New Horizons spacecraft an Annual Check-Out (ACO), with observations relevant to this work (i.e. observations of specific star clusters) taken every other year. Details of the observations used for the radiometric calibration of the Red, Blue, NIR, CH4, Pan 1, Pan 2 and Pan Frame cameras are given in Tables 1 to 4. As they show most of the early observations (2008 to 2012) were of the Messier 6 and 7 clusters (NGC 6405 and 6475, or sometimes shortened to M6 and M7), while later observations (2013 to 2014) also included observations of the Wishing Well Cluster (NGC 3532). This change was made to include a larger number and variety of star types, to help with both the geometric distortion correction and radiometric calibration. An example of a typical image is shown in Figure 2. The 5.7 degree field of view of MVIC is large enough to capture both the M6 and M7 clusters in a single image, which allows many stars to be observed simultaneously.

*2.1 Overview of the Modeling Technique*

The software written to perform this calibration was developed by many people, over many years. The basic premise of the software is to compare the flux observed by MVIC of a given star with an expected model flux. If MVIC were perfectly calibrated the two would be identical, while any offset between them is the calibration offset this work seeks to determine. The offset for each of MVIC's detectors has to be separately determined.

*2.2 Modeling the Stellar Flux*

The first task is to find the stars in the MVIC field of view. Once this task is achieved the next steps are to determine which stars they are and then calculate the photon flux expected from each star (to be compared eventually to the one observed). These first two steps are by far the most complicated aspect of the model, as it is possible to easily mistake hot-pixels as stars and miscorrelate stars with those cataloged.

The software finds all the potential stars in a given MVIC image by searching for bright pixels above a 5 data-number (DN) threshold value. It then uses the pointing information in the header to determine the Right Ascension (RA) and Declination (Dec) for each of these sources. The positions of all the stars are then compared to those in the Tycho-2 star catalog (Hog *et al.* 2000a, 200b), while those that are missing are assumed to either simply be missing from the Tycho catalog or are false-positives (for example cosmic ray strikes). The catalog provides star's spectral type, Tycho V and B (referred to henceforth as $B_T$ and $V_T$ respectively) magnitudes, and temperature. From these values the Johnson V ($V_J$) magnitude is determined for each star from the Tycho magnitudes, according to Hog et al. (2000c): $V_J = V_T - 0.09\ (B_T - V_T)$.

The Kurucz 1993 Atlas (Kurucz, 1993) is used to determine the expected emission from each star using the Tycho catalog's stellar temperature and by assuming a solar abundance. The Kurucz model best able to fit these two requirements is used to give each star's emission across the full wavelength range of each filter ($F_\lambda$), and at 5556 Å ($F_{5556Å}$). The stellar emission at each wavelength is then scaled to an absolute emission ($F_S(\lambda)$) using Vega as the standard star, to account for the different distances of the target stars. Recently Bohlin *et al.* (2014) determined the absolute emission of Vega at 5556 Å to be 3.44e-9 erg s$^{-1}$ cm$^{-2}$ Å$^{-1}$, Vega is defined to be 0 magnitude in the $V_J$ band, so the absolute emission of each star at wavelength is given by Equation 3.

$$F_S(\lambda) = \frac{F_\lambda}{(F_{5556Å})(3.44e-9)\left(10^{(V_J/-2.5)}\right)} \qquad [Equation\ 3]$$

This absolute energy flux is then converted to a number flux ($F_n$), which describes the number of photons emitted at each wavelength using the relationship $F_n(\lambda) = F_s(\lambda) / \left(hc/\lambda\right)$, where $h$ is the Planck constant and $c$ is the speed of light. The final step is to take this number flux and determine how many of these photons are able to hit the MVIC detector. This final flux, $F_e$, gives the expected count rate (the number of photons per second hitting the detector) from a given star. It is calculated as a function of the detector response ($\chi$) and the aperture size of the circular detector (3.75 cm² radius, Reuter *et al.*, 2008) according to $F_e = (\pi\, 3.75^2) \int F_n(\lambda)\chi(\lambda)d\lambda$. This flux can now be directly compared to the flux observed.

*2.3 Star photometry*

For each star, we perform basic aperture photometry to measure the total flux and its associated uncertainty (in counts) of both the star and the surrounding sky using the IDL routine basphote.pro (Buie, 2015). This is achieved by using the image's exposure time, and by assuming a read-out noise of 30 electrons, a gain of 58.6 electrons/photon, a sky annulus between 10 and 20 pixels, and an aperture size of 4 pixels (Reuter *et al.*, 2008). The results from this routine were checked against other standard photometry algorithms (e.g. aper.pro see Buie, 2015) and the results were found to be consistent.

*2.3 Stellar Calibration Results*

Figure 3 compares the observed and model count rate for all observations made with each of the MVIC detectors. The line of best fit to the data (shown in red) provides the required adjustment factor (to go from observed to actual count rates). These values for each of the detectors are listed in Table 5, along with the error of the mean. The close agreement of the model and observed star counts implies that correlate stars is correctly identifying the stars in the images.

## 3 The Effect of the Electronics Side on MVIC images

On approach to Pluto it was observed that the gain of the NIR channel drifted if the output was read through the primary electronics side (known as side 1), whilst it remained stable when read through the backup electronics side (side 0). During New Horizons' cruise to the Pluto system MVIC's annual checkout observations were alternated between electronic sides but this effect was not discovered, primarily because there were too few observations taken on each electronic side to make finding robust trends possible. This problem with the NIR channel was not discovered during New Horizons' 2007 encounter with Jupiter because all science observations were taken using the same electronic side (side 1). However, upon approach to the Pluto system observations were made using altering electronics sides, and by systematic inspection of Charon's color in the NIR the gain drift became apparent.

Thus, it was important to check whether this gain fluctuation affected the adjustment factors required to move from observed to model count rates for all the MVIC detectors. Figures 4 and 5 show the model versus observed count rates for observations made only with side 0 or side 1. These results are summarized in Figure 6 and Table 5, which show that most of the detectors do not show a significant change in adjustment factor with electronic side. The close agreement of the adjustment values of the three Pan detectors in Figure 6 provides confidence that the derived values are correct, since they cover the same wavelength range but have been independently derived. The two detectors that show notable change in their adjustment values with electronic sides are the NIR and Pan 1 detector. This problem with Pan 1 was first discovered back during the instrument's commissioning phase in 2006, which was early enough to be able to mitigate the problem by ensuring all science observations were taken using Pan 1's good electronics side (side 1). However, because the problem with the NIR channel was not known until shortly before encounter similar mitigation steps were not enacted for it. Therefore, post-observation processing solutions had to be used instead, as described in the next section.

Figure 6 also shows that there is significant difference in the adjustment factors between detectors. The Blue channel requires negligible adjustment (1.00±0.01), whilst the NIR and CH4 channels require a ~20 to 45% correction respectively, with the other channels lying between these extremes. It is unclear why these two channels require the most correction. Although it is worth noting that the wavelength range of the CH4 filter (860-910 nm) is much narrower than the other filters (see Figure 1), so fewer stars are observed its images (see Table 5) leading to higher errors in its adjustment value.

Since the cause of the problems with NIR and Pan 1 channels are not understood it was possible they could vary over time. Therefore, the adjustment factor was determined for each filter for each year, as shown in Figure 7. This was possible because although all

Pan 1 science observations were made using the electronic side 1 annual checkout observations were made on both sides to monitor the problem. Figure 7 shows that for all detectors (except the two detectors that are known to problematic: NIR and Pan 1) that the adjustment factors agree within error every year. Furthermore, there are no obvious temporal dependence in the NIR and Pan 1 adjustment factors, and their deviations from the average adjustment factor is not increasing with time (if anything the opposite is true for the NIR detector).

**4 Charon Calibration Process**

We also derived a channel-wise relative radiometric calibration from Charon observations for the Blue, NIR, and CH4 channels, scaled from the Red channel stellar calibrations. The disk-averaged color ratios of Charon were matched to those that would be produced by the product of a parametric synthetic reflection spectrum (pinned to a mean F555W geometric albedo of 0.41) and a solar spectrum that also reproduces the global color ratios of Charon as measured by the Hubble Space Telescope's (HST) ACS HRC in the F435W and F555W filters (Buie *et al.*, 2006).

Since Charon is broadly characterized by two latitudinally-controlled color units (neutral mid-latitudes and a red polar cap), the relative contributions of these two color units was adjusted in order to match the orientation of Charon as observed by HST between 2002 and 2003. We found that this geometric correction to the global color ratios of Charon was negligible. Note, the details of C_COLOR2 and all other observations discussed in this section and the ones that follow are given in Table 7.

The parametric reflection spectrum that we adopted in order to reproduce the Buie et al. (2006) Charon colors had the form

$$p(\lambda) = \left(1 + e^{0.495 - 2.43 \times 10^{-4}\lambda}\right)^{-1}, \qquad [\textit{Equation 4}]$$

with $\lambda$ in nm.

Correction factors relative to the Red channel were determined for Blue, NIR, and CH4 channel transmission curves as follows:

1) The ratio of the global mean flux of C_COLOR2 images, with regional contributions reweighted to correct them to HST's viewing geometry, were calculated for Red, Blue, NIR, and CH4 channels.
2) The observed ratios of Blue, NIR, and CH4 to Red were determined.
3) The expected ratios of Blue, NIR, and CH4 to Red were determined by integrating the product of the solar spectrum, the parametric Charon reflection spectrum (Equation 4), and each filter's transmission curve over wavelength.
4) The ratio of the expected ratio over the observed ratio was adopted as the *relative* throughput correction factor for Blue, NIR, and CH4. These and the Red channel were all scaled by the *absolute* correction factor (Table 1) derived for the Red channel from the stellar calibrations.

## 5 Charon Calibration Results

The throughput curves used in this analysis before and after correction, as well as the parametric Charon spectrum, are illustrated in Figure 8. The Charon-based calibrations produce results very similar to the stellar calibrations, demonstrating a cross-validation of the two unique approaches. The largest variation between the two calibration solutions was in the CH4 channel, for which the Charon-based calibration determined a correction factor 3-7% larger (depending on power side) than the stellar calibrations. This channel has the pivot wavelength farthest from the pivot wavelengths of the two HST filters used to calibrate the parametric Charon reflectance spectrum, and it is not surprising that this is where the largest difference between the two calibration approaches appeared.

## 6 Bootstrapping a Power Side Correction

The calibration solution for images affected by this gain drift includes the derivation of a correction for affected channels using repeated imagery of the same terrains. This "bootstrapping" process has proven to be very effective for the flyby data, as the Pluto system contains two excellent calibration targets. The first is Charon, which has little to no longitudinal color variations and provides an excellent "gray card" for approach imagery where both Pluto and Charon are visible in a single MVIC FOV. The second is informally called Sputnik Planum, which is a large and flat region that is extremely uniform in color and albedo and which is visible at high resolution in the imagery where Charon is not available.

For a given image set affected by gain drift, the following process is used to derive a bootstrap correction: (1) identify the temporally-nearest image set not affected by gain drift that contains overlapping imagery of either Sputnik Planum or Charon; (2) co-register the imagery on a map grid; (3) extract pixels from a contiguous region of equal surface area from both image sets within either Sputnik Planum or the disk of Charon; (3) determine the summed flux within these pixels in the Red channel and the channel of interest *i*; (4) compute the ratio of these summed fluxes between the channel of interest *i* and the Red channel in both image sets; and (5) compute the ratio of these two ratios, which is the correction factor:

$$CF_i = \left( \left( \sum F_{i,SIDE\ 0} \bigg/ \sum F_{Red,SIDE\ 0} \right) \bigg/ \left( \sum F_{i,SIDE\ 1} \bigg/ \sum F_{Red,SIDE\ 1} \right) \right) \qquad [Equation\ 5]$$

For consistency, all channels except Red (which serves as our control channel and does not appear to be affected by gain drift) are corrected in this way, though the derived NIR correction is always substantially larger than those derived for Blue or CH4. What follows is a worked example for P_COLOR_2, which was taken on side 1 of the electronics and was therefore subject to gain drift. In this example we chose PC_MULTI_MAP_B17 as the control imagery, as it was taken on side 0 of the electronics (which is not subject to gain drift) and it covered a similar sub-spacecraft longitude and shows Sputnik Planum clearly.

The images were extracted to an interim common map projection, and a circular region (on the sphere) with a radius of 10º was extracted from the core of Sputnik Planum at approximately 20º North, 180º East in both images. In this region, the median raw Red/NIR DN ratio for P_COLOR_2 was 0.766, while for PC_MULTI_MAP_B17 it was 0.815; the ratio of these two determines the bootstrapped NIR gain correction factor for P_COLOR_2, 1.064. This correction factor was found to be insensitive to whether ratios were determined from mean color ratios, median color ratios, or the ratios of areal sums (the latter being adopted). For P_COLOR_2, derived gain correction factors for Blue (1.039) and CH4 (1.019) were substantially smaller.

To test for robustness, the maps were intentionally misregistered by up to 5 degrees North and South, the extracted radius shrunk to 5º, and the bootstrapping process repeated. This keeps the region of interest within the boundaries of Sputnik Planum, but incorrectly correlates different regions within Sputnik Planum. Due to the uniformity of Sputnik Planum's colors on large spatial scales, the derived correction ratios varied by only small amounts (~1%), demonstrating that the corrections are robust to small misregistration between image sets if Sputnik Planum is used as a control region. Charon provides similar robustness due to its longitudinal color uniformity.

## 7 Radiometric Calibration Keywords

To transform DN detected by MVIC into physical units describing the incoming spectral energy distribution, MVIC image headers contain two calibration-dependent keywords.

For the first Planetary Data System (PDS) release, these keywords are defined based on the stellar calibrations described in this document. The diffuse source sensitivity keywords are defined as:

$$R_{TARGET,FILTER} = \int_0^\infty \frac{S_{TARGET}(\lambda)\chi_{TARGET}(\lambda)A\lambda\Theta}{S_{TARGET}(\lambda_{p,FILTER})hc\,e^-/DN}d\lambda,$$

while the point-source sensitivity keywords are defined as

$$P_{TARGET,FILTER} = \int_0^\infty \frac{S_{TARGET}(\lambda)\chi_{TARGET}(\lambda)A\lambda}{S_{TARGET}(\lambda_{p,FILTER})hc\,e^-/DN}d\lambda.$$

Where $S_{TARGET}(\lambda)$ is a source-dependent spectrum in $erg\,s^{-1}cm^{-2}A^{-1}$ defined at a surface one AU from the target, $\chi_{FILTER}(\lambda)$ is the responsivity, which is the same as previously described except it uses the filter transmission curves for a specific filter after throughput is corrected by the adjustment factors from Table 6, $\lambda_{p,FILTER}$ is the pivot wavelength of a specific filter, $A$ is the aperture collecting area in $cm^2$ for the MVIC telescope, $\frac{e^-}{DN}$ is the mean MVIC gain, and $\Theta$ is the MVIC pixel IFOV in steradians (19.77 μrad by 19.77 μrad, Reuter *et al.*, 2008). These keywords are defined for SOLAR, PLUTO, CHARON, JUPITER, and PHOLUS target spectra. The spectra used to define these keywords are derived from the following sources: Charon (Buie and Grundy, 2000), Pluto (Douté *et al.,* 1999), Pholus (Cruikshank *et al.,* 1998) and Solar (Colina *et al.,* 1996). The actual spectra used will be delivered to the PDS as part of the next delivery by the New Horizons project; some have been slightly updated.

# 8 Conclusion

We have described the two semi-independent methods used to calibrate New Horizons' MVIC instrument. The close agreement between the two methods provides some reassurance that both are functioning correctly. The "Charon" calibration was used to make science products widely throughout New Horizons' encounter with Pluto. However, the stellar calibration will be used in future PDS deliveries primarily because it was produced for all color filters and does not rely on having a known color target in the field of view. We also describe a previously known problem with the Pan 1 filter, and the observational strategy adopted to minimize its effect. Finally we have also described the newly discovered gain problem with the NIR detector and the bootstrapping technique that has been adopted to mitigate it.

# 9 Tables

| Array Name | Array Description | Wavelength Range (nm) | Pivot Wavelength (nm) | Array Size (pixels) |
|---|---|---|---|---|
| Pan 1 | Panchromatic TDI #1 | 400 - 975 | 692 | 5024x32 |
| Pan 2 | Panchromatic TDI #2 | 400 - 975 | 692 | 5024x32 |
| Blue | Blue TDI | 400-550 | 492 | 5024x32 |
| Red | Red TDI | 540-700 | 624 | 5024x32 |
| NIR | Near-Infrared TDI | 780-975 | 861 | 5024x32 |
| CH4 | Methane-Band TDI | 860-910 | 883 | 5024x32 |
| Pan Frame | Panchromatic Framing Camera | 400-975 | 692 | 5024x128 |

Table 1: Details of the MVIC arrays. A single pixel is 19.77 μrad by 19.77 μrad, so the FOV of the TDI array is 5.7° by 0.037°, and that of the framing camera is 5.7° by 0.146°.

| Mid-Observation Time (UTC) | Onboard Mission Elapsed Time (MET) | Right Ascension (°) | Declination (°) | Exposure Time (s) | Target |
|---|---|---|---|---|---|
| 2008-10-15T04:45:25.191 | 0086351808 | 266.842 | -33.457 | 2.919 | NGC 6405 and 6475 |
| 2010-06-25T21:30:25.129 | 0139807309 | 266.767 | -33.431 | 2.853 | NGC 6405 and 6475 |
| 2012-06-01T21:15:25.784 | 0200891209 | 266.771 | -33.434 | 2.880 | NGC 6405 and 6475 |
| 2013-07-03T06:45:27.183 | 0235139809 | 166.433 | -58.754 | 2.811 | NGC 3532 |
| 2014-07-22T13:50:25.578 | 0268342909 | 266.859 | -33.480 | 2.863 | NGC 6405 and 6475 |
| 2014-07-22T13:57:19.578 | 0268343327 | 266.775 | -33.371 | 4.244 | NGC 6405 and 6475 |
| 2014-07-22T18:00:27.079 | 0268357909 | 166.433 | -58.786 | 2.762 | NGC 3532 |

Table 1 – Details of the stellar observations used to calibrate the MVIC color channels (Red, Blue, NIR, CH4). All channels observed simultaneously.

| Mid-Observation Time (UTC) | Onboard Mission Elapsed Time (MET) | Right Ascension (°) | Declination (°) | Exposure Time (s) | Target |
|---|---|---|---|---|---|
| 2008-10-15T05:00:11.691 | 0086352708 | 266.731 | -33.471 | 2.891 | NGC 6405 and 6475 |
| 2010-06-25T21:49:11.629 | 0139808449 | 266.780 | -33.528 | 2.802 | NGC 6405 and 6475 |
| 2012-06-01T21:32:11.284 | 0200892228 | 266.708 | -33.461 | 2.828 | NGC 6405 and 6475 |
| 2013-07-03T07:12:15.183 | 0235141429 | 166.391 | -58.710 | 2.901 | NGC 3532 |
| 2014-07-22T14:10:11.578 | 0268344108 | 266.732 | -33.477 | 2.862 | NGC 6405 and 6475 |
| 2014-07-22T18:15:14.579 | 0268358809 | 166.487 | -58.685 | 2.762 | NGC 3532 |

Table 2 – Details of the stellar observations made by Panchromatic Filter #1, used in this calibration.

| Mid-Observation Time (UTC) | Onboard Mission Elapsed Time (MET) | Right Ascension (°) | Declination (°) | Exposure Time (s) | Target |
|---|---|---|---|---|---|
| 2008-10-15T04:52:11.691 | 086352228 | 266.764 | -33.509 | 2.845 | NGC 6405 and 6475 |
| 2010-06-25T21:40:11.629 | 0139807909 | 266.717 | -33.505 | 2.766 | NGC 6405 and 6475 |
| 2012-06-01T21:24:12.284 | 0200891749 | 266.753 | -33.501 | 2.842 | NGC 6405 and 6475 |
| 2013-07-03T07:03:15.183 | 0235140889 | 166.516 | -58.730 | 2.902 | NGC 3532 |
| 2014-07-22T18:07:15.079 | 0268358329 | 166.491 | -58.720 | 2.850 | NGC 3532 |

Table 3 – Details of the stellar observations made by Panchromatic Filter #2, used in this calibration.

| Mid-Observation Time (UTC) | Onboard Mission Elapsed Time (MET) | Right Ascension (°) | Declination (°) | Exposure Time (s) | Target |
|---|---|---|---|---|---|
| 2008-10-15T23:55:11.126 | 0086420815 | 346.135 | -7.185 | 0.500 | Neptune |
| 2010-06-25T21:57:10.238 | 0139808935 | 266.761 | -33.531 | 1.000 | NGC 6405 and 6475 |
| 2012-06-01T21:40:09.893 | 0200892714 | 266.838 | -33.461 | 1.000 | NGC 6405 and 6475 |
| 2012-06-02T01:28:10.393 | 0200906396 | 270.405 | -14.638 | 0.500 | Pluto |
| 2012-06-02T01:28:34.663 | 0200906425 | 270.431 | -14.636 | 1.000 | Pluto |
| 2013-07-02T19:30:10.291 | 0235099315 | 266.692 | -33.506 | 1.000 | NGC 6405 and 6475 |
| 2014-07-23T17:23:05.535 | 0268442094 | 270.696 | -14.635 | 1.000 | Pluto |
| 2014-07-23T17:15:06.535 | 0268441615 | 270.350 | -14.444 | 1.000 | Pluto |
| 2014-07-22T14:16:09.687 | 0268344474 | 266.785 | -33.562 | 1.000 | NGC 6405 and 6475 |

Table 4 – Details of the stellar observations made by Panchromatic Frame Camera, used in this calibration.

| Filter | Both AF | Side 0 AF | Side 1 AF | Both #Stars | Side 0 #Stars | Side 1 #Stars | Charon AF |
|---|---|---|---|---|---|---|---|
| Red | 1.21±0.01 | 1.23±0.01 | 1.21±0.01 | 621 | 269 | 352 | 1.21* |
| Blue | 1.00±0.01 | 1.02±0.01 | 0.99±0.01 | 324 | 149 | 175 | 1.02 |
| NIR | 1.32±0.01 | 1.38±0.02 | 1.27±0.01 | 405 | 143 | 262 | 1.39 |
| CH4 | 1.46±0.02 | 1.44±0.04 | 1.51±0.03 | 102 | 23 | 79 | 1.56 |
| Pan 1 | 1.17±0.01 | 1.27±0.01 | 1.13±0.01 | 636 | 157 | 479 | NA |
| Pan 2 | 1.22±0.01 | 1.23±0.01 | 1.21±0.01 | 528 | 313 | 215 | NA |
| Pan Frame | 1.26±0.01 | 1.28±0.01 | 1.23±0.01 | 668 | 228 | 440 | NA |

Table 5 – List of the adjustment factors (AF) required to correct the observed to expected counts rate, where the expected count rate = (observed count rate x adjustment factor) using all the stars observed. *Charon-based Red channel adjustment factor is set by the stellar calibration.

|  | Adjustment Factor | | | Number of Stars | | |
| --- | --- | --- | --- | --- | --- | --- |
| Year | All | Side 0 | Side 1 | All | Side 0 | Side 1 |
| **Red** | | | | | | |
| 2008 | 1.23±0.01 | 1.23±0.01 | - | 61 | 61 | - |
| 2010 | 1.20±0.02 | - | 1.20±0.02 | 98 | - | 98 |
| 2012 | 1.21±0.02 | 1.21±0.02 | - | 77 | 77 | - |
| 2013 | 1.21±0.01 | - | 1.21±0.01 | 113 | - | 113 |
| 2014 | 1.21±0.01 | 1.24±0.01 | 1.21±0.01 | 272 | 131 | 141 |
| **Blue** | | | | | | |
| 2008 | 0.99±0.01 | 0.99±0.01 | - | 36 | 36 | - |
| 2010 | 0.97±0.01 | - | 0.97±0.01 | 47 | - | 47 |
| 2012 | 1.00±0.01 | 1.00±0.01 | - | 41 | 41 | - |
| 2013 | 1.01±0.01 | - | 1.01±0.01 | 51 | - | 51 |
| 2014 | 1.02±0.01 | 1.05±0.01 | 1.00±0.02 | 149 | 72 | 77 |
| **NIR** | | | | | | |
| 2008 | 1.46±0.03 | 1.46±0.03 | - | 40 | 40 | - |
| 2010 | 1.18±0.02 | - | 1.18±0.02 | 86 | - | 86 |
| 2012 | 1.46±0.06 | 1.46±0.06 | - | 42 | 42 | - |
| 2013 | 1.42±0.03 | - | 1.42±0.03 | 52 | - | 52 |
| 2014 | 1.30±0.01 | 1.34±0.02 | - | 185 | 61 | 124 |
| **CH4** | | | | | | |
| 2008 | 1.45±0.05 | 1.45±0.05 | - | 12 | 12 | - |
| 2010 | 1.48±0.07 | - | 1.48±0.07 | 17 | - | 17 |
| 2012 | 1.42±0.06 | 1.42±0.06 | - | 11 | 11 | - |
| 2013 | 1.53±0.05 | - | 1.53±0.05 | 17 | - | 17 |
| 2014 | 1.51±0.04 | - | 1.51±0.04 | 45 | - | 45 |
| **Pan 1** | | | | | | |
| 2008 | 1.28±0.02 | 1.28±0.02 | - | 91 | 91 | - |
| 2010 | 1.25±0.02 | - | 1.25±0.02 | 70 | - | 70 |
| 2012 | 1.25±0.02 | 1.25±0.02 | - | 66 | 66 | - |
| 2013 | 1.28±0.01 | - | 1.28±0.01 | 147 | - | 147 |
| 2014 | 1.02±0.01 | - | 1.02±0.01 | 262 | - | 262 |
| **Pan 2** | | | | | | |
| 2008 | 1.24±0.02 | 1.24±0.02 | - | 97 | 97 | - |
| 2010 | 1.20±0.02 | - | 1.20±0.02 | 83 | - | 83 |
| 2012 | 1.26±0.02 | 1.26±0.02 | - | 86 | 86 | 0 |
| 2013 | 1.23±0.01 | - | 1.23±0.01 | 132 | - | 132 |
| 2014 | 1.18±0.01 | 1.18±0.01 | - | 130 | 130 | - |
| **Pan Frame** | | | | | | |
| 2008 | 1.31±0.01 | 1.31±0.01 | - | 22 | 22 | - |
| 2010 | 1.22±0.02 | - | 1.22±0.02 | 124 | - | 124 |
| 2012 | 1.28±0.01 | 1.28±0.01 | - | 206 | 206 | - |
| 2013 | 1.20±0.01 | - | 1.20±0.01 | 129 | - | 129 |
| 2014 | 1.32±0.01 | - | 1.32±0.01 | 187 | - | 187 |

Table 6 - List of the adjustment factors required to correct the observed to expected counts rate (see Table 5 for more details) using all stars observed each year.

| Observation Name | Target | Mid-Observation Time (UTC) | Onboard Mission Elapsed Time (MET) | Phase (degrees) | Electronic Side |
|---|---|---|---|---|---|
| PC_MULTI_MAP_B17 | Pluto and Charon | 13-July-2015 03:38:06 | 0299064592 | Pluto:15.6 Charon:15.5 | 0 |
| C_COLOR2 | Charon | 14-Jul-2015 10:42:28 | 0299176432 | 38.6 | 0 |
| P_COLOR_2 | Pluto | 14-Jul-2015 11:10:52 | 0299178092 | 38.8 | 1 |

Table 7 - Details of the observations used in the "Charon" calibration.

# 10 Figures

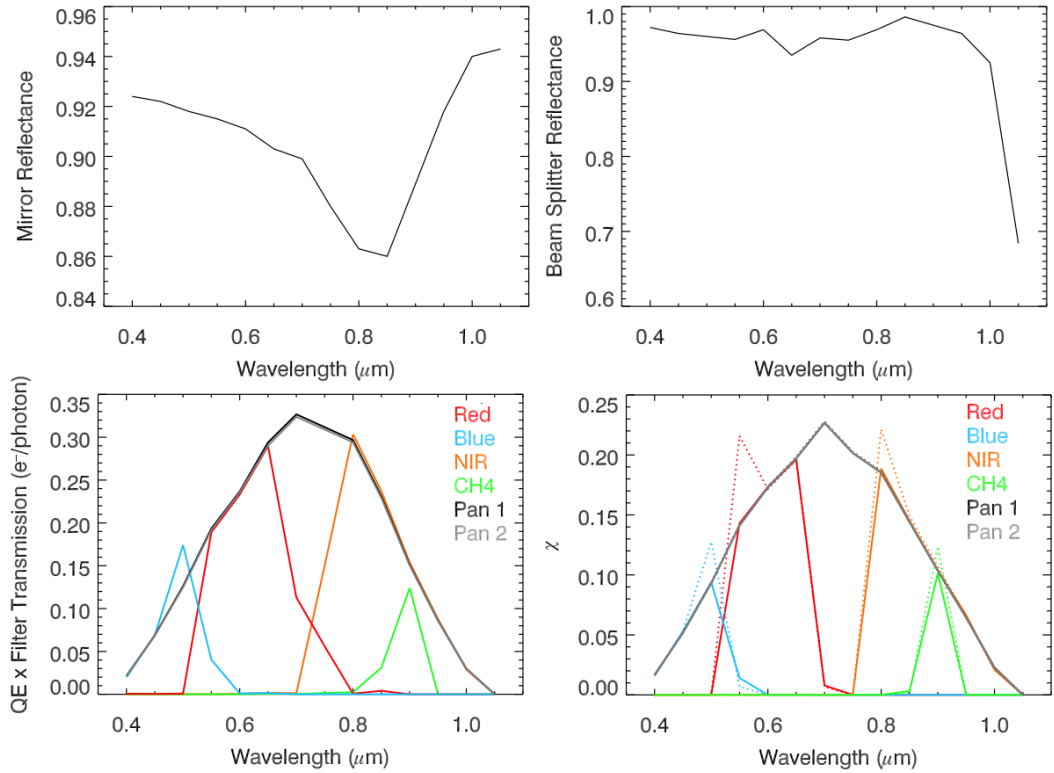

Figure 1 – Mirror and beamsplitter reflectance, the product of the quantum efficiency ($Q_e$), and the different filter transmission curves are shown. These combine, as described in the main text, to produce the responsivity ($\chi$). Both the original responsivity (dotted lines) and tweaked responsivity (solid lines) are shown. See main text for details.

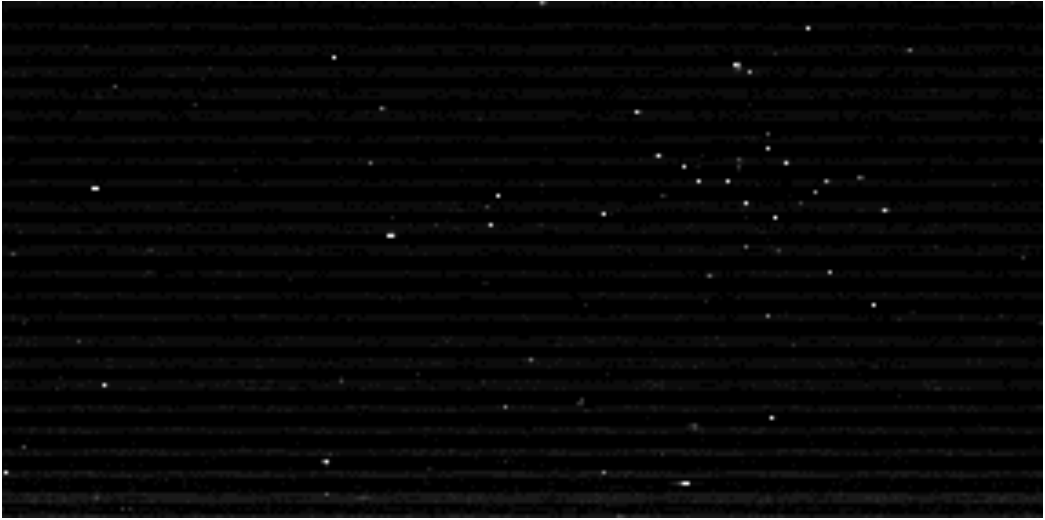

Figure 2 – MVIC Red image from the M6/M7 cluster observations taken in 2015 (MET 0268342909).

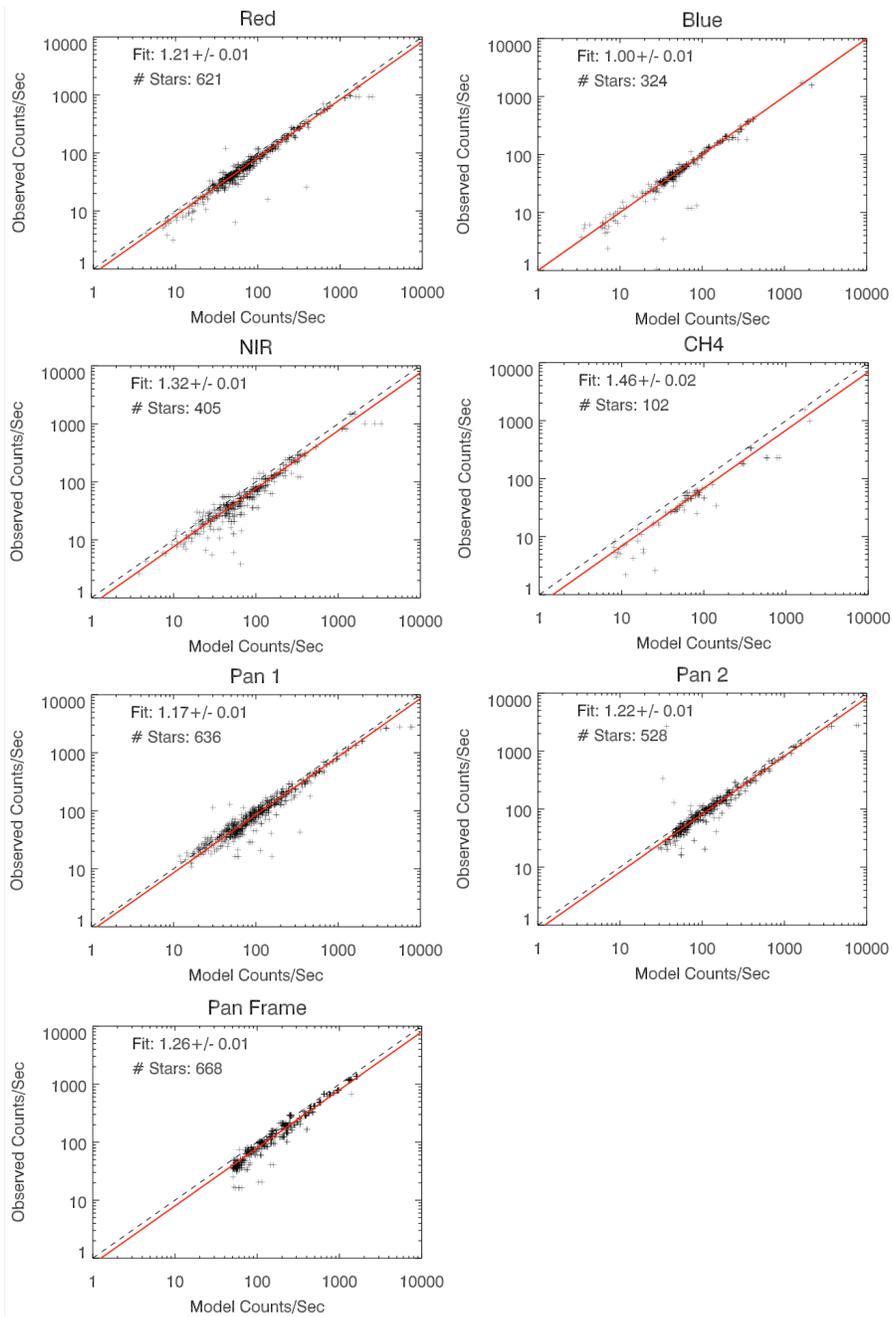

Figure 3 – Plots of the model versus the observed counts/second for all observations made with each MVIC filter. The black dotted line shows x=y, and the red solid line shows the best fit, as given each of the figures and Table 5. The total number of stars (crosses) is also given in each of the figures.

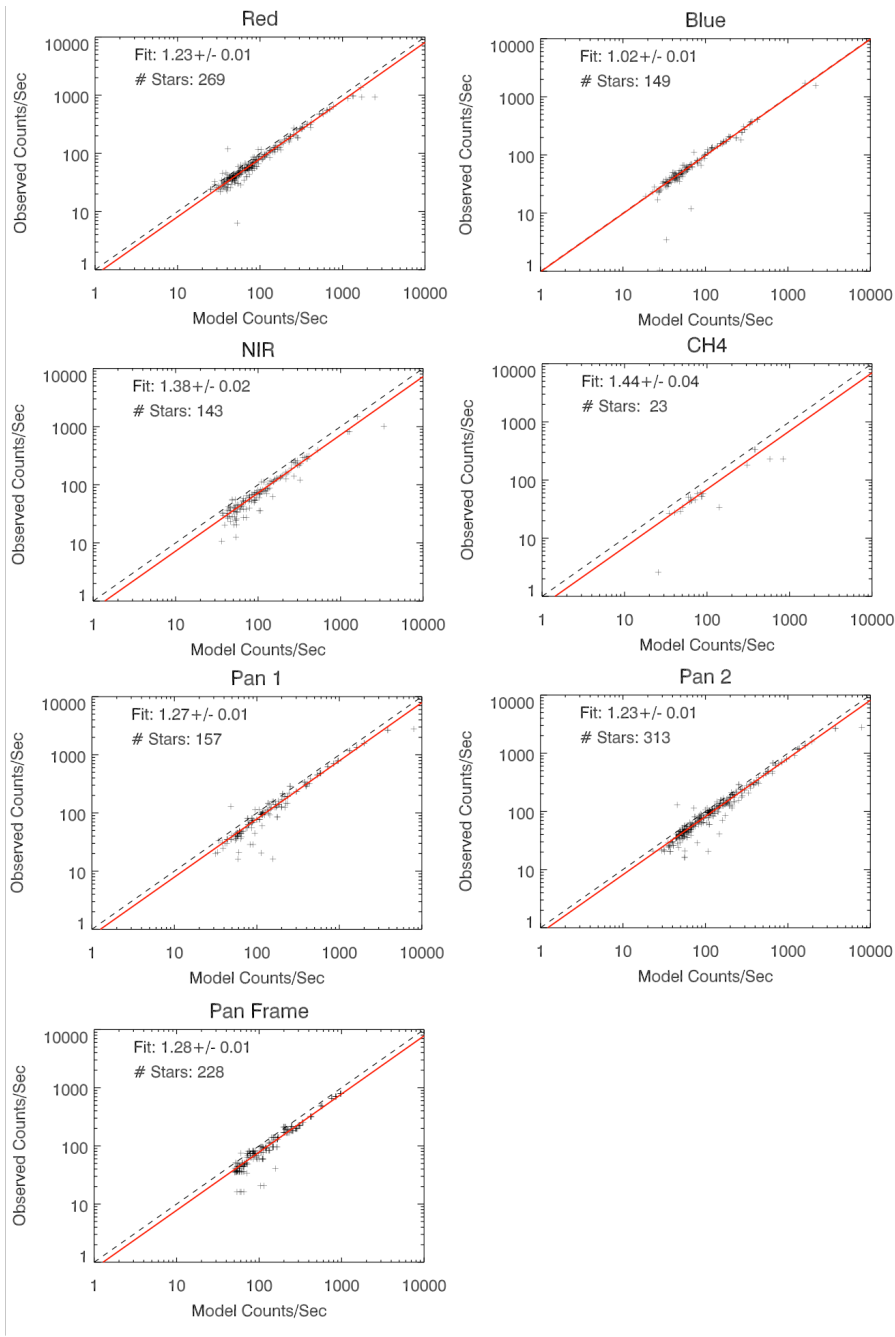

Figure 4 – Plots of the model versus the observed counts/second for all observations made on the electronics side 0 with each MVIC filter. The black dotted line shows x=y, and the red solid line shows the best fit, as given each of the figures and Table 5. The total number of stars (crosses) is also given in each of the figures.

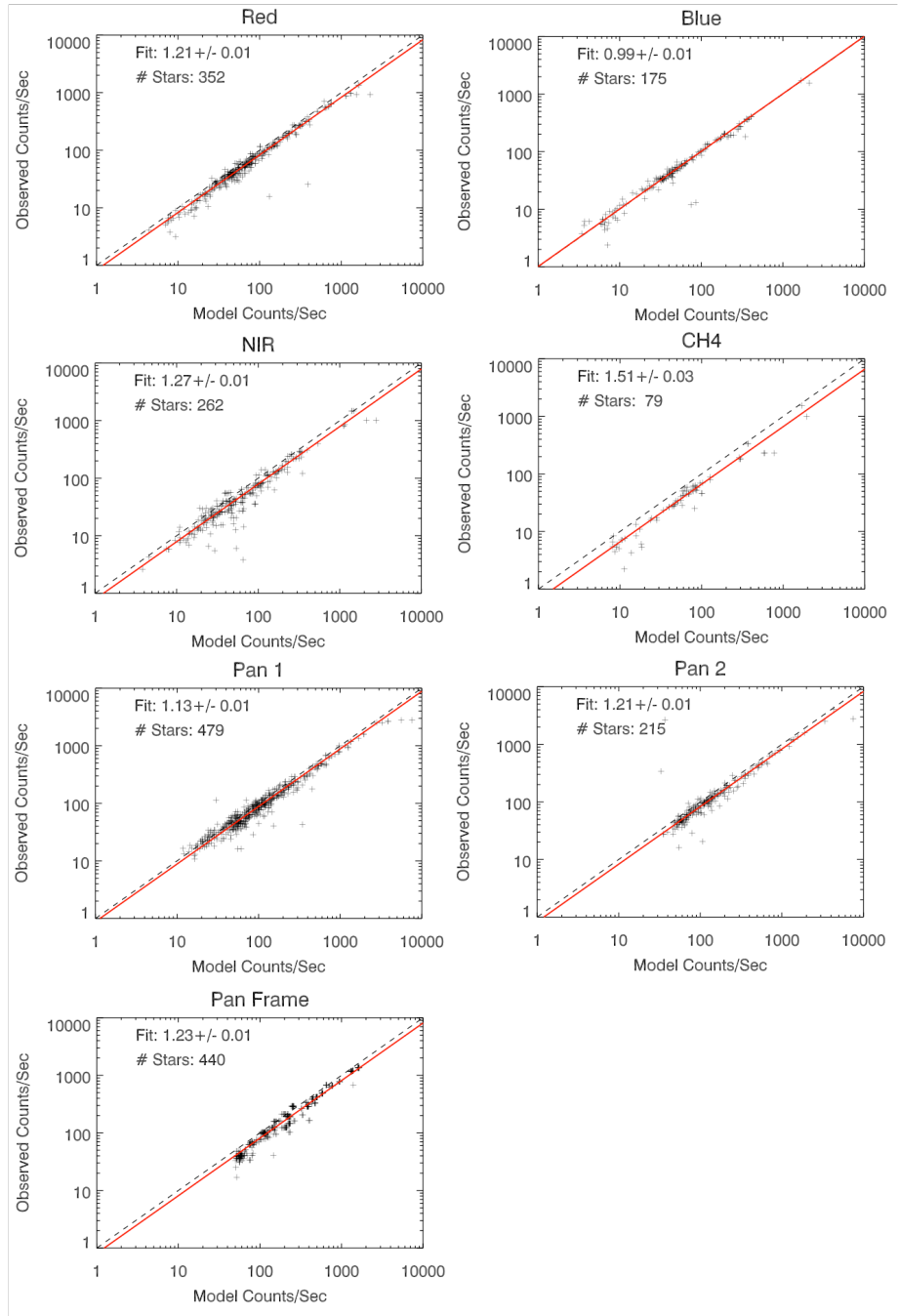

Figure 5 – Plots of the model versus the observed counts/second for all observations made on the electronics side 1 with each MVIC filter. The black dotted line shows x=y, and the red solid line shows the best fit, as given each of the figures and Table 5. The total number of stars (crosses) is also given in each of the figures.

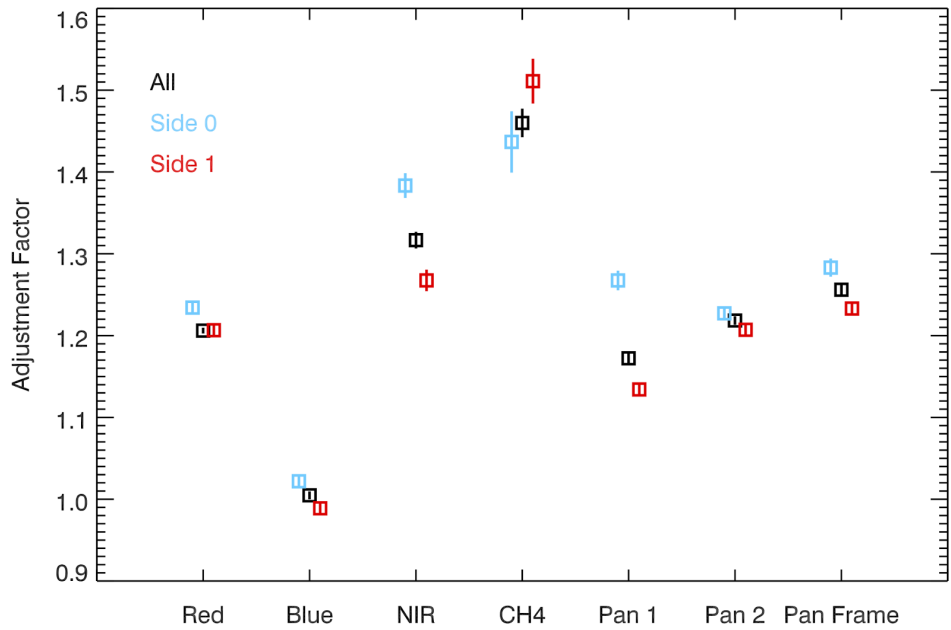

Figure 6 – The adjustment ratio of all MVIC filters, as calculated for all electronic sides, just side 0 and just side 1.

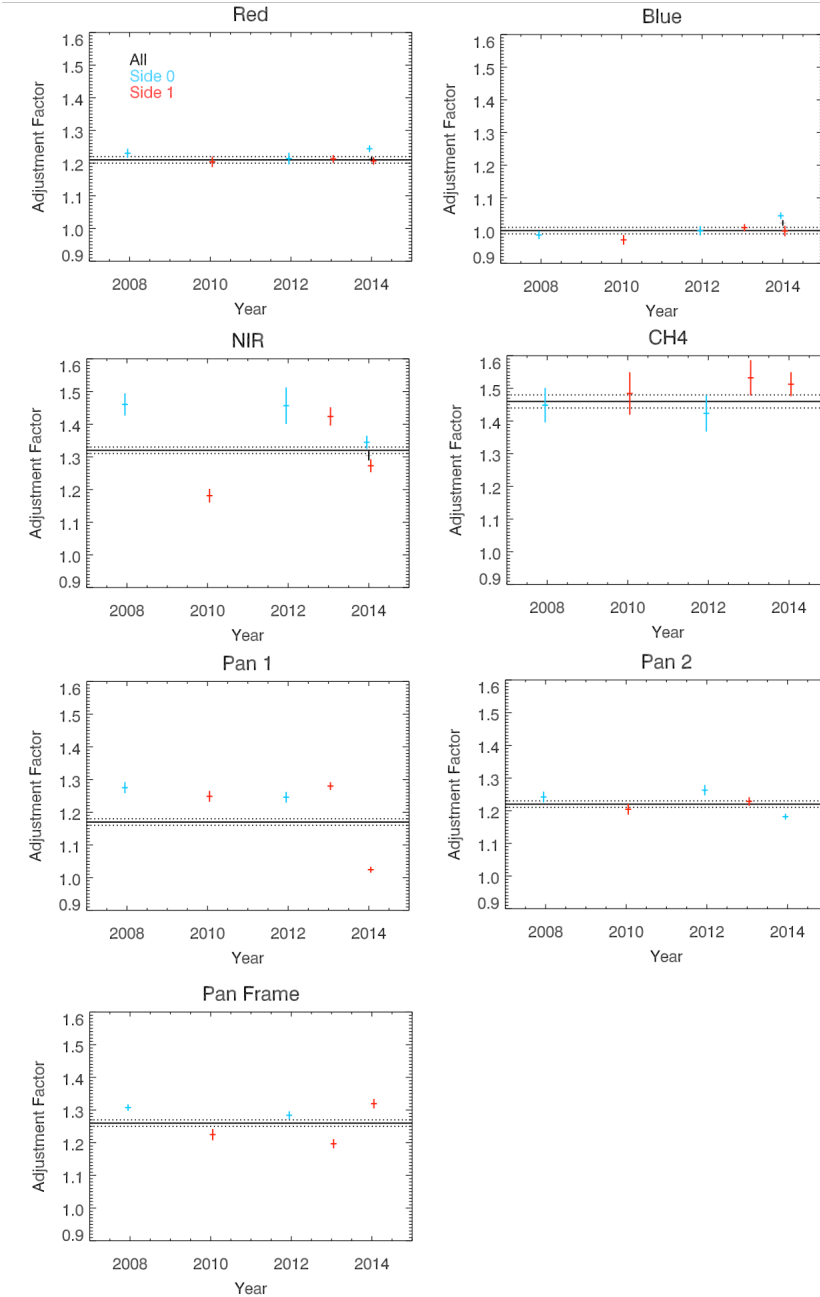

Figure 7- The adjustment ratio of all the MVIC filters plotted by year, as calculated for all electronic sides, just side 0 and just side 1 where available. The black solid and dotted lines shows the adjustment value and the error of the mean for each filter, which was derived from all available data as given in Table 5.

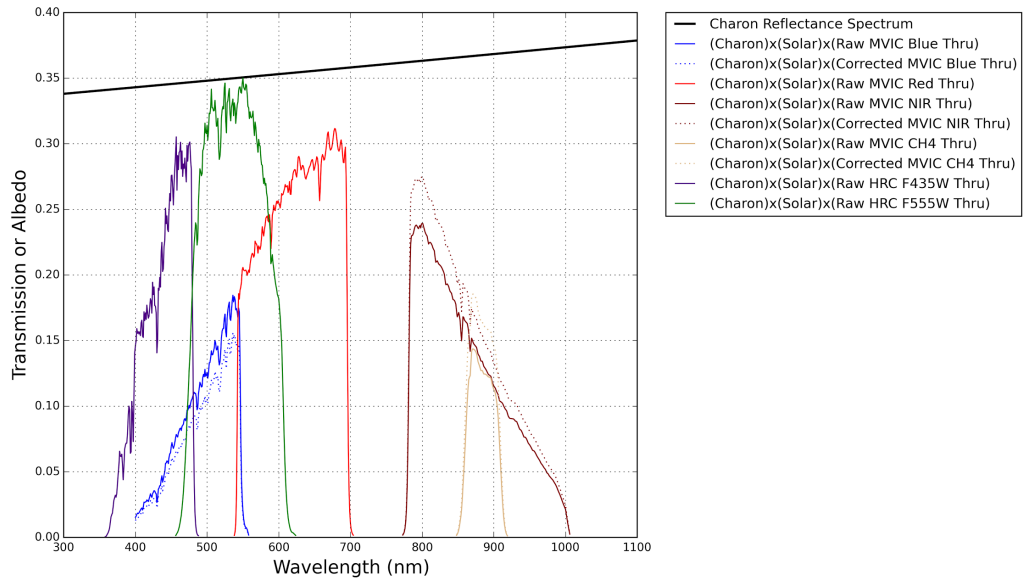

Figure 8- Raw input and calibrated transmission curves (scaled by product of solar spectrum and model Charon spectrum) for Charon-based calibration procedure, and derived parametric Charon reflectance spectrum (geometric albedo, black line). HST ACS HRC F435W and F555W filters accessed from http://www.stsci.edu/hst/acs/analysis/throughputs on October 29, 2015.

## 11 Acknowledgements

This work was supported by NASA's New Horizons Project. S. Philippe and B. Schmitt acknowledge the Centre National d'Etudes Spatiales (CNES) for its financial support through its "Système Solaire" program.